\documentclass[%
reprint,
superscriptaddress,
amsmath,amssymb,
aps
]{revtex4-2}

\usepackage{graphicx}
\usepackage[dvipsnames]{xcolor}
\usepackage{dcolumn}% Align table columns on decimal point
\usepackage{bm}% bold math
\usepackage[utf8]{inputenc}
\usepackage[T1]{fontenc}
\usepackage{amsmath}

\draft % marks overfull lines with a black rule on the right

\begin{document}
%\preprint{APS/123-QED}
% Use the \preprint command to place your local institutional report number 
% on the title page in preprint mode.
% Multiple \preprint commands are allowed.
%\preprint{}

\title[Dong et al.]{Spin-orbit coupling in digital alloyed InGaAs quantum wells}

\author{Jason T. Dong}
%\email[]{jtdong@ucsb.edu}
%\homepage[]{Your web page}
%\thanks{}
\altaffiliation{Present Address: Laboratory for Physical Sciences, 8050 Greenmead Drive, College Park, 20742, USA}
\affiliation{Materials Department, University of California, Santa Barbara, California 93106, USA}

\author{Yilmaz Gul}
%\email[]{Your e-mail address}
%\homepage[]{Your web page}
%\thanks{}
%\altaffiliation{}
\affiliation{London Centre for Nanotechnology, University College London, 17-19 Gordon Street, London WC1H
0AH, United Kingdom}

\author{Irene Villar Rodriguez}
%\email[]{Your e-mail address}
%\homepage[]{Your web page}
%\thanks{}
%\altaffiliation{}
\affiliation{London Centre for Nanotechnology, University College London, 17-19 Gordon Street, London WC1H
0AH, United Kingdom}

\author{Aaron N. Engel}
%\email[]{Your e-mail address}
%\homepage[]{Your web page}
%\thanks{}
%\altaffiliation{}
\affiliation{Materials Department, University of California, Santa Barbara, California 93106, USA}

\author{Connor P. Dempsey}
%\email[]{Your e-mail address}
%\homepage[]{Your web page}
%\thanks{}
%\altaffiliation{}
\affiliation{Department of Electrical and Computer Engineering, University of California, Santa Barbara, California 93106, USA}

\author{Stuart N. Holmes}
%\email[]{Your e-mail address}
%\homepage[]{Your web page}
%\thanks{}
%\altaffiliation{}
\affiliation{Department of Electronic and Electrical Engineering, University College London, Torrington Place,
London WC1E 7JE, United Kingdom}

\author{Michael Pepper}
%\email[]{Your e-mail address}
%\homepage[]{Your web page}
%\thanks{}
%\altaffiliation{}
\affiliation{London Centre for Nanotechnology, University College London, 17-19 Gordon Street, London WC1H
0AH, United Kingdom}
\affiliation{Department of Electronic and Electrical Engineering, University College London, Torrington Place,
London WC1E 7JE, United Kingdom}

\author{Christopher J. Palmstr\o m}
\email[]{cjpalm@ucsb.edu}
%\homepage[]{Your web page}
%\thanks{}
%\altaffiliation{}
\affiliation{Materials Department, University of California, Santa Barbara, California 93106, USA}
\affiliation{Department of Electrical and Computer Engineering, University of California, Santa Barbara, California 93106, USA}

\date{\today}

\begin{abstract}
Increasing the spin-orbit coupling in InGaAs quantum wells is desirable for applications involving spintronics and topological quantum computing. Digital alloying is an approach towards growing ternary quantum wells that enables asymmetric interfaces and compositional grading in the quantum well, which can potentially modify the spin-orbit coupling in the quantum well. The spin-orbit coupling of the quantum wells is extracted from beating patterns in the low magnetic field magnetoresistance. Digital alloying is found to modify the spin-orbit coupling by up to 138 meV\textnormal{\AA}. The changes induced in the spin-orbit coupling can be qualitatively understood as being due to modifications in the interfacial Rashba spin-orbit coupling.

\end{abstract}

% insert suggested PACS numbers in braces on next line

\maketitle %\maketitle must follow title, authors, abstract and \pacs

\section{Introduction}

Spin-orbit coupling (SOC) is a relativistic effect that can cause normally spin degenerate bands to become spin split due to inversion symmetry breaking. InGaAs quantum wells (QW) with strong spin-orbit coupling are potentially useful for spintronics \cite{Datta1990,Chuang2015}, topological quantum computing \cite{Oreg2010,Lutchyn2010}, and as a platform for mesoscopic physics experiments \cite{Liu2023,Delfanazari2024}. Increasing the spin-orbit coupling of the quantum well is often desirable. In spin field effect transistors, a stronger spin orbit coupling enables shorter channel length and thus allows for better scaling. In the application of topological quantum computing, increasing the spin-orbit coupling is expected to increase the topological gap of the system. The reported topological gaps are $\sim10$ $\mu$eV \cite{PhysRevB.107.245423}, and the parameter space the topological gap is reported to exist in is extremely narrow, making the scaling of multiple devices challenging. Increasing the spin-orbit coupling is a potential approach to address these problems. 

In III-V quantum wells such as InGaAs, Dresselhaus \cite{Dresselhaus1955} and Rashba \cite{Rashba1960} spin-orbit coupling are both present. Dresselhaus spin-orbit coupling arises from the inversion asymmetry of the crystal structure, while Rashba spin-orbit coupling arises from inversion asymmetry of the heterostructure due to interfaces and electric fields within the heterostructure. The Dresselhaus spin splitting is typically significantly smaller than the Rashba spin splitting in III-V quantum wells \cite{Ganichev2004,Giglberger2007,Knox2018,Farzaneh2024}. The spin-orbit coupling in the heterostructures can be increased with the Rashba effect. By increasing the electric field within the quantum well, typically with a gate electrode, the Rashba spin-orbit coupling can be increased. Increasing the spin-orbit coupling of the system with an external electrical field can cause detrimental effects in the heterostructure. At larger electric fields within the quantum well, the mobility of the system can decrease due to the quantum confined Stark effect increasing interface roughness scattering \cite{Jana2011}, and the second subband can become occupied, which can also reduce mobility due to intersubband scattering \cite{Wickramasinghe2018,Zhang2023,Farzaneh2024}. It is desirable to increase the spin-orbit coupling of the material system without significantly reducing the charge carrier mobility of the system to take advantage of the potential benefits high spin-orbit coupling can offer.

% The linear Rashba spin-orbit coupling parameter ($\alpha$) can be separated into two components, an electric field depend component ($\alpha_{F_E}$) and a component due to the interface ($\alpha_i$).

Digital alloying is a potential approach towards increasing the spin-orbit coupling of the system. Digital alloys (DA) are short period superlattices that can approximate the properties of random alloys. Enhancement of the charge carrier mobility by digital alloying was recently demonstrated \cite{Dong2024prm}. Digital alloying can allow for increased asymmetry in the heterostructure by being able to form asymmetric interfaces at the top and bottom interface of the quantum well as well as enabling an easy method of compositional grading within the heterostructure by varying the relative sublattice thicknesses. Due to these properties of digital alloys, it is anticipated that this technique may be a promising method to grow asymmetric quantum wells and increase the spin-orbit coupling of the system. To date, negligible changes in the spin-orbit coupling in triangular GaAs quantum wells have been demonstrated by digital alloying \cite{Eldridge2011}, and it has been predicted that digital alloying can induce modest changes to the Rashba spin-orbit coupling in triangular InGaAs and InSb quantum wells \cite{Pingenot2011}.

\begin{figure*}[t!]
\centering
\includegraphics[width=\textwidth]{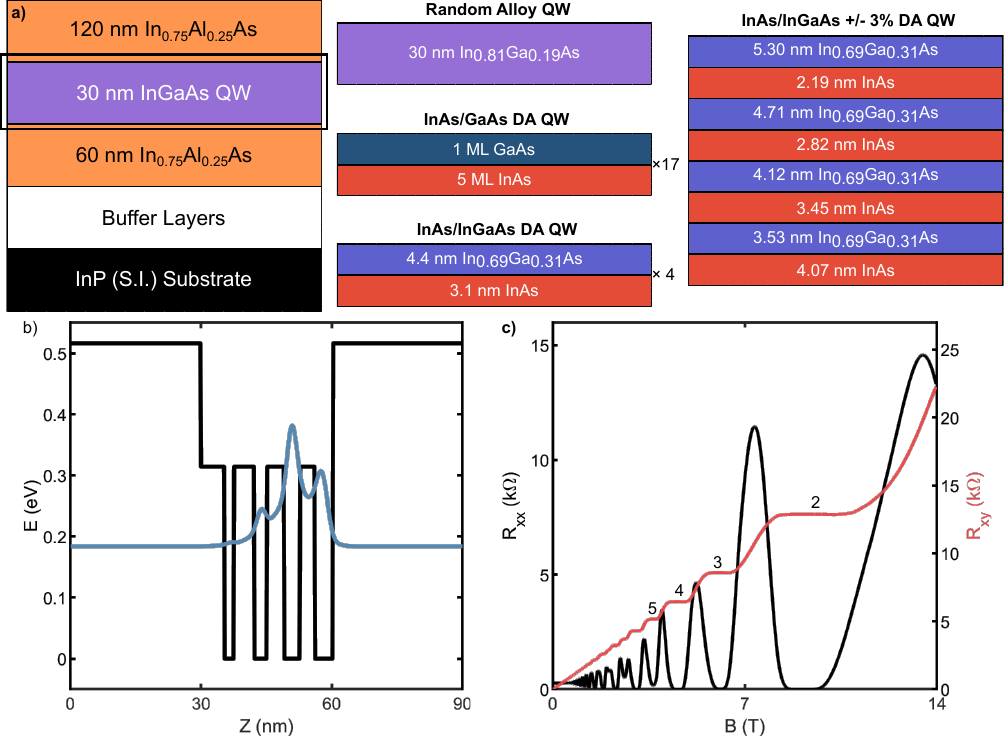}
\caption{\label{fig:soLayers} a) Schematic layer structure consisting of a 30 nm InGaAs quantum well buried 120 nm from the surface on the left. Layer structure of the different quantum wells, consisting of a random alloy quantum well, a InAs/GaAs DA quantum well, a InAs/InGaAs DA quantum well, and a 3\% graded InAs/InGaAs DA quantum well. b) Conduction band edge and $k \cdot p$ calculations of the probability density of the 3\% graded InAs/InGaAs DA quantum well. c) 2 K magnetoresistance of the 3\% graded InAs/InGaAs DA quantum well.}
\end{figure*}

In this work, the effect of digital alloying on the spin-orbit coupling in In\textsubscript{0.81}Ga\textsubscript{0.19}As QW grown on InP substrates is investigated. A random alloy QW, a nongraded InAs/GaAs DA QW, a nongraded InAs/InGaAs DA QW, and a 3\% compositionally graded InAs/InGaAs DA were studied. Significant increases in the Rashba coefficient were observed in the InAs/GaAs DA sample, while more modest changes were observed in the graded and nongraded InAs/InGaAs DA sample. The changes in both of the InAs/InGaAs DA samples can qualitatively be explained with $k \cdot p$ theory, while the changes in the InAs/GaAs DA sample require further investigation. The Rashba coefficient in graded DA sample is also found to be more responsive to the gate voltage. These results are indicative that digital alloying is a promising approach towards engineering the spin-orbit coupling in semiconductor quantum wells.

\section{Methods}

The samples consisted of a 30 nm In\textsubscript{0.81}Ga\textsubscript{0.19}As quantum well with an 120 nm thick In\textsubscript{0.75}Al\textsubscript{0.25}As top barrier grown on a metamophic InAlAs virtual substrate on semi-insulating InP (001) wafers with solid source molecular beam epitaxy in a VG V80H MBE system. The exact details of the growth are given in \cite{Dong2024prm}. The quantum wells were either a random alloy QW, a InAs/GaAs DA QW, a InAs/InGaAs DA QW, and a 3\% graded InAs/InGaAs DA QW. The InAs/GaAs DA consisted of a 17 period superlattice of 1 monlayer of InAs and 5 monolayers of GaAs. The nongraded InAs/InGaAs DA QW consisted of a 4 period superlattice of InAs and In\textsubscript{0.69}Ga\textsubscript{0.31}As. The 3\% compositional grading was implemented by varying the relative thickness of InAs and InGaAs but maintaining a total period thickness of 7.5 nm. The exact layer structure is given in Fig. \ref{fig:soLayers} a. In all of the DA QWs, the bottom interface of the QW is an InAs/InAlAs interface, while the top interface of the QW is an (In)GaAs/InAlAs interface.

Hall bars were fabricated to study the electrical transport properties. The Hall bar mesa was etched with wet chemical etching using a mixture of H\textsubscript{2}SO\textsubscript{4}:H\textsubscript{2}O\textsubscript{2}:H\textsubscript{2}O (1:8:120) along the [110] direction. Ohmic contacts to the quantum well were formed with NiGeAu contacts  annealed at 450 \textsuperscript{o}C for 2 minutes. To modulate the carrier density in the Hall bar, a 30 nm AlO\textsubscript{x} gate dielectric was deposited with atomic layer deposition followed by a Ti/Au top gate electrode. The width of the Hall bar was 80 $\mu$m, with an arm spacing of 740 $\mu$m. The ratio of the Hall bar width to the spacing of the arms was 9.25. Unless stated otherwise, transport measurements were performed at 2 K. Temperature dependent measurements were performed with a modulation field, using the experimental methods described in \cite{Holmes2008}.

\section{Results and discussion}

\subsection{Properties of a compositionally graded quantum well}
The predicted band structure with an 8 band $k \cdot p$ model \cite{Vurgaftman2021} and the magnetotransport of the graded InAs/InGaAs DA QW is shown in Fig. \ref{fig:soLayers} b \& c. The band structure and the magnetotransport of the other samples are given in \cite{Dong2024prm}. From the probability density of the 1\textsuperscript{st} electron conduction band subband, the wavefunction of the QW is asymmetric due to the compositional grading implemented by the digital alloying and shifted towards the bottom interface. There are peaks of probability density in the InAs regions of the quantum well and troughs in the InGaAs regions. The magnetotransport of the 3\% graded QW shows the integer quantum Hall effect with vanishing magnetoresistance, indicative of a high quality 2DEG without any parallel conduction. The onset of SdH oscillations occurs at B = 0.22 T. The integer quantum Hall effect (IQHE) also indicates that the DA QW is not a multiple QW. In multiple QWs the IQHE gains additional quantization by the number of multiple QWs \cite{Stormer1986}, the lack of additional quantization in the DA samples indicates the sample is not a multiple QW. The peak mobility of the graded DA QW is 524,000 cm\textsuperscript{2}/Vs at a carrier density of $4.13 \times 10^{11}$ cm\textsuperscript{-2}, which is lower than the peak mobility of 545,000 cm\textsuperscript{2}/Vs in the nongraded quantum well \cite{Dong2024prm}. The lower mobility of this sample compared to the nongraded DA QW in these heterostructures is due to increased interface roughness scattering due to the closer proximity of the peak in the probability density to interfaces.

\subsection{Gate control of spin splitting in a quantum well}

\begin{figure}[]
\centering
\includegraphics[width=3.375in]{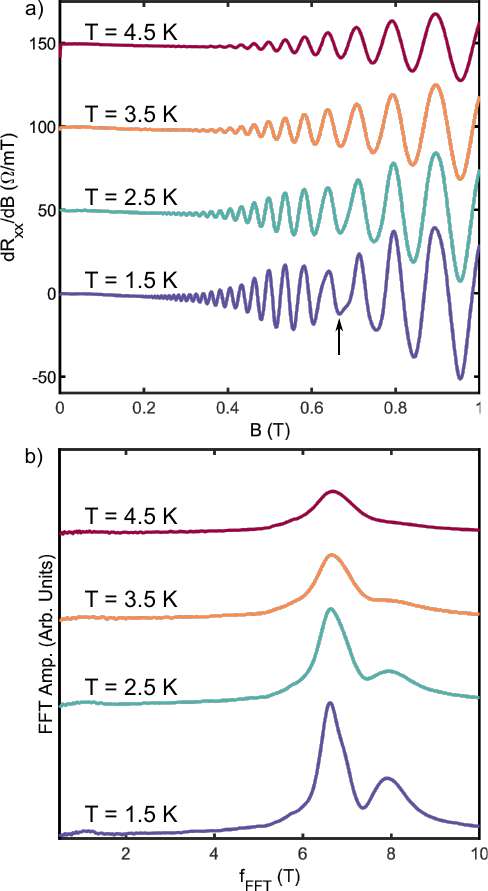}
\caption{\label{fig:soTFFT} a) Temperature dependent dR\textsubscript{xx}/dB of the InAs/GaAs DA QW.  b) FFT of the dR/dB oscillations. At low temperatures a clear peak splitting due to spin orbit-coupling can be observed, at higher temperatures the oscillations thermalize and the peak splitting disappears.}
\end{figure}

\begin{figure*}[]
\centering
\includegraphics[width=\textwidth]{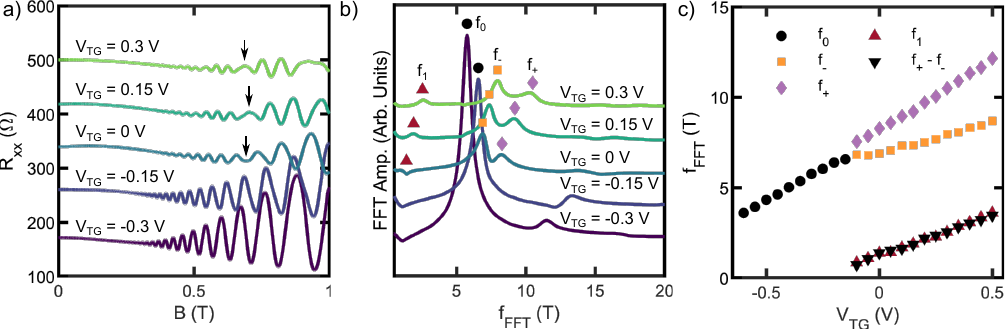}
\caption{\label{fig:soRxxVg} a) Gate voltage dependence of the InAs/GaAs DA QW magnetoresistance. A beating pattern is observed at larger gate voltages. b) FFT of the magnetoresistance, showing the peak in the FFT spectra spin split into two peaks at larger gate voltages. A low frequency MIS peak can be observed once the spin splitting occurs, corresponding to the frequency difference between the spin split peaks. c) Top gate voltage dependence of the spin degenerate and spin split frequencies.}
\end{figure*}

Due to the high mobility of the material system, a beating pattern in the magnetoresistance can be observed at low magnetic fields (Fig. \ref{fig:soTFFT}a). In the Fourier transform of the oscillations in 1/B, two peaks in the Fourier transform can be observed, see Fig. \ref{fig:soTFFT}b. Beating patterns in the magnetoresistance can be due to SdH oscillations of zero magnetic field spin splitting, magnetointersubband scattering (MIS) \cite{Raikh1994,Sander1998}, and magnetophonon resonances (MPR) \cite{Gurevich1961,Tsui1980,Chen2018}. MIS generates a frequency in the magnetoresistance that corresponds the difference in frequency ($\Delta$f) between the oscillations of two bands. The beating patterns generated by MIS and MPR are relatively insensitive to the measurement temperature. It is expected the beating patterns generated by spin-orbit coupling have a strong temperature dependence, since the beating patterns should possess the same temperature dependence as SdH oscillations. To rule out MIS and MPR the origin of the beating pattern, the temperature dependence of the beating pattern was measured (Fig. \ref{fig:soTFFT}). From the temperature dependence, the beating pattern rapidly disappears with temperature, indicating the origin of this beating pattern is SdH oscillations.

The beating pattern in the low field magnetotransport data is attributed to zero-field splitting of the Fermi surface due to spin-orbit coupling. In the case of linear Rashba and Dresselhaus spin-orbit coupling, the splitting energy is parameterized by:
$\Delta_{so} = 2 \alpha k_f $, where $\alpha$ is the spin-orbit coupling coefficient and $k_f$ is the Fermi wavevector. Both Rashba and Dresselhaus spin-orbit coupling cause the Fermi surface to split into two spin-split Fermi surfaces. These spin split Fermi-surfaces do not possess the same surface area and thus are occupied by different carrier densities. The different carrier densities of the Fermi surface cause the beating patterns observed in magnetotransport. The dependence of the beating pattern with the applied gate voltage is shown in Figs. \ref{fig:soRxxVg}a \& b, where the gate voltage dependence of the InAs/GaAs DA QW magnetoresistance and the Fourier analysis of the magnetoresistance is shown. It can be observed that at lower gate voltages, a single SdH oscillation is observed, corresponding to a single peak in the Fourier transform spectra (f\textsubscript{0}). As the gate voltage increases, the frequency of the oscillation increases, consistent with the carrier density of the quantum well increasing, and the Rashba spin-orbit coupling increases. Eventually, the Rashba spin-orbit coupling and the spin-splitting become large enough enough to be observed as a beating pattern in the magnetoresistance and f\textsubscript{0} is observed to split into two peaks (f\textsubscript{-} \& f\textsubscript{+}). A low frequency peak now appears (f\textsubscript{1}), and the frequency of f\textsubscript{1} is in good agreement with the $\Delta$f between f\textsubscript{-} \& f\textsubscript{+}. The change in frequencies of the peaks at different gate voltages and the splitting of f\textsubscript{0} for the InAs/GaAs DA quantum well is shown in Fig. \ref{fig:soRxxVg}c. For all of the applied gate voltages, the frequency of the f\textsubscript{1} is in good agreement with the difference in frequency of the two spin-split peaks. This behavior is consistent with f\textsubscript{1} peak being due to MIS between the two spin-split peaks, similar to what has been observed before in HgTe and InGaAs quantum wells \cite{Minkov2020}. As the gate voltage increases, the spin split peaks shift to higher frequencies, and the frequency difference between the two spin-split peaks increases. In all of the InGaAs QW samples, similar gate control over the spin splitting is demonstrated, which indicates that the Rashba spin-orbit coupling is the dominant spin-orbit coupling mechanism in these quantum wells.

\subsection{Influence of digital alloying on the spin-orbit coupling}

\begin{figure}[]
\centering
\includegraphics[width=3.375in]{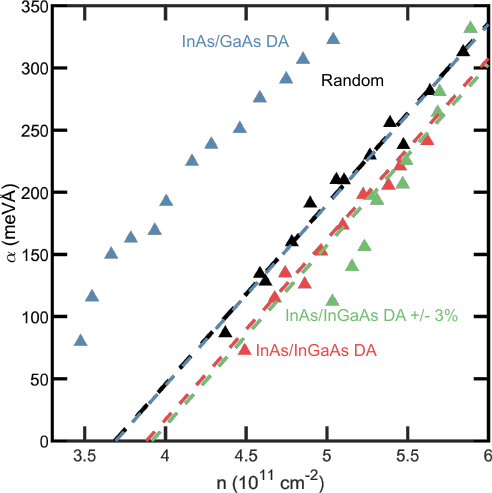}
\caption{\label{fig:soRashba} SOC parameter at different carrier densities for the different samples. The carrier density here is extracted from the Shubnikov-de Haas oscillations. The black line is a linear fit of the SOC parameter to the random alloy QW, with the colored lines offset from the linear fit to the random alloy QW by the predicted differences in the interface Rashba SOC contribution.}
\end{figure}

The linear spin-orbit coupling parameter can be extracted from the frequencies of the spin splitting. For a spin split band, the frequency can be converted into a carrier density with the following formula $n^{\pm} = ef_{\pm}/h$, where $e$ is the elementary charge and $h$ is the Planck constant. With these frequencies, the linear spin-orbit coupling parameter can now be determined \cite{Engels1997}:
\begin{equation}
    \alpha = \frac{(n^{+} - n^{-}) \hbar^2}{m^*} \sqrt{\frac{\pi}{2(n^+ + n^-)- 2(n^+ - n^-)}}
\end{equation}
The SOC parameter at different carrier densities, with the total carrier density determined from SdH oscillations, is shown in Fig. \ref{fig:soRashba}. The SOC parameter increases with increases to the gate voltage and thus the carrier density. The gate tunability of the SOC parameter is consistent with a change in the SOC parameter due to electric field dependent contribution from Rashba SOC. At lower carrier densities, the random alloy QW SOC parameter is similar what has been previously reported in InGaAs QW \cite{Holmes2008,Chen2015}. The InAs/GaAs DA QW has higher SOC parameter than the random alloy QW, while both of the InAs/InGaAs DA QW samples have a lower SOC parameter than the random alloy QW at most carrier densities. At higher carrier densities, the SOC parameter of the 3\% graded InAs/InGaAs DA QW exceeds that of the random alloy QW.

With $k \cdot p$ theory, the Rashba SOC parameter ($\alpha_R$) can be separated into an electric field component ($\alpha_{{E}}$) and an interface component ($\alpha_{{i}}$) \cite{Engels1997, Schapers1998}. 
\begin{equation}
    \alpha_{R} = \alpha_{{E}} + \alpha_{{i}}
\end{equation}
The $\alpha_{{i}}$ can be explicitly calculated with the equation \cite{Engels1997, Schapers1998,Eldridge2011}: 
\begin{eqnarray}
\alpha_{i} = \frac{\hbar^2 E_{p} \hat{z}}{6 m_0} \sum_{n} \bigg(\frac{\Delta \Gamma_{7}}{2(E-\Gamma_{7}^{+} - \phi(z_{n}))^2}\nonumber \\
+ \frac{\Delta \Gamma_{7}}{2(E-\Gamma_{7}^{-} - \phi(z_{n}))^2} - \frac{\Delta \Gamma_{8}}{2(E-\Gamma_{8}^{+} - \phi(z_{n}))^2}\nonumber \\
- \frac{\Delta \Gamma_{8}}{2(E-\Gamma_{8}^{-} - \phi(z_{n}))^2}\bigg)|\Psi(z_{n})|^2 
\label{eq:RashbaI}
\end{eqnarray}
Where $\hat{z}$ is the unit vector, $E_{p}$ is the Kane energy, $\Gamma_{7}$ is the split-off band edge, $\Gamma_{8}$ is the valance band edge, $\phi$ is the electrostatic potential, $|\Psi|^2$ is the probability density, $E$ is the energy, $n$ is the index for every interface in the heterostructure, $\Delta \Gamma_{7}$ is the split-off band offset of at the interface, $\Delta \Gamma_{8}$ is the valence band offset of at the interface, and $\pm$ is an index for the top and bottom material at the interface. $|\Psi|^2$ is calculated using an 8 band $k \cdot p$ model, with the details of the model and the parameters used for the calculation of $|\Psi|^2$ and $\alpha_{{i}}$ provided in \cite{Vurgaftman2021}. The $\alpha_{{i}}$ exhibits a dependence on the electric field within the samples due to the increased asymmetry of the wavefunction induced by larger electric fields. 
% move to appendix?

Both of the nongraded DA QW have similar slopes to the change of the SOC parameter with carrier density as the random alloy QW, and their dependence appears to be offset from the dependence of the SOC parameter with carrier density of the random alloy QW. The InAs/InGaAs DA QW has a 20 meV\textnormal{\AA} smaller SOC parameter, while the InAs/GaAs DA QW has a 138 meV\textnormal{\AA} larger SOC parameter. The similar slopes are indicative that the change in the Rashba SOC due to the electric field from the gate is the same, but the nongraded DA QW has a different zero electric field $\alpha_{{i}}$. However, the 3\% graded InAs/InGaAs DA QW appears to have a much steeper slope, and the SOC parameter appears to change more rapidly with changes in carrier density. The steeper slope is likely due to the $\alpha_{{i}}$ having a stronger electric field dependence from the greater asymmetry in the wavefunction due to compositional grading.

\begin{table}[b]
\caption{\label{tab:1}
The change in predicted $\alpha_{{i}}$ from that of the random alloy calculated with $k \cdot p$ theory.}
\begin{ruledtabular}
\begin{tabular}{ l c c c}
  
 Sample & InAs/GaAs & InAs/InGaAs & InAs/GaAs 3\% \\ 
\hline
$\Delta \alpha_{i}$ (meV\textnormal{\AA})& -0.8 & -28.3 & -33.4 \\
\end{tabular}
\end{ruledtabular}
\end{table}

The linear fit to the random alloy QW SOC parameter was offset by the calculated difference in $\alpha_{{i}}$ at zero electric field between the DA QW and the random alloy QW, shown in Fig. \ref{fig:soRashba}. The $\alpha_{{i}}$ was calculated using equation \ref{eq:RashbaI}. The change in predicted $\alpha_{{i}}$ from that of the random alloy is given in Table \ref{tab:1}. There is good agreement between the measured SOC parameter of the nongraded InAs/InGaAs DA QW and the fit to the random alloy QW SOC parameter offset by the zero electric field $\alpha_{{i}}$, which is indicative that the difference is due to the interface contribution to the Rashba effect. The compositional grading in the 3\% graded InAs/InGaAs DA QW is predicted to induce a larger change $\alpha_{{i}}$, and the larger decrease in observed in the SOC parameter of this sample at lower carrier densities qualitatively agrees with this prediction. The overall decrease in the total SOC of the graded and nongraded InAs/InGaAs DA QW can be explained as being due the direction of the change in Rashba SOC induced from the digital alloying being opposite to the direction of the Rashba SOC induced from the gate electrode. As a result, there is a net decrease in the total SOC of the InAs/InGaAs DA QW. These results show that the changes in the spin-orbit coupling can be qualitatively explained by the interface contribution to the Rashba effect. However, both the direction and the larger magnitude of the change in SOC in the InAs/GaAs DA QW cannot be explained with $k \cdot p$ theory, and this change likely requires further theoretical investigation to understand.

% If the quantum well heterostructure were to be inverted, with the bottom interface with the barrier being InGaAs and the top interface with the barrier being InAs, it is expected that the directions of the interface and electric field effects to align, and a overall enhancement of the total spin-orbit coupling to be observed.

\section{Conclusions}

In summary, digital alloying was demonstrated to significantly modify the SOC of the quantum wells in InGaAs quantum wells. Gate control of the SOC in the quantum wells was observed for the random alloy and the digital alloy quantum wells. Changes up to 138 meV\textnormal{\AA} of the SOC parameter were observed, with the largest changes in the InAs/GaAs DA QW. Nongraded DA QW were observed to have similar changes in SOC parameter with gate voltage, but are offset from each other. The changes in the nongraded DA samples can be qualitatively understood as being due to changes in the Rashba parameter induced from differences at the interfaces. The compositionally graded DA QW demonstrated stronger gate tunability of the SOC in the quantum well, which is likely due to the increased asymmetry of the quantum well from the compositional grading. Further theoretical investigation is required to quantitatively explain the changes in the SOC. These results establish digital alloying as an effective method of modifying the SOC of quantum wells.

\section{Acknowledgements}
The growth of the quantum wells was supported by University of California Multiple Campus Award No. 00023195. Transport experiments were supported by the Department of Energy under award No. DE-SC0019274 and UK Science and Technology Facilities Council under award No. ST/Y005074/1. The use of the Nanotech UCSB Nanofabrication Facility is acknowledged.

\bibliography{References}

%apsrev4-2.bst 2019-01-14 (MD) hand-edited version of apsrev4-1.bst
%Control: key (0)
%Control: author (8) initials jnrlst
%Control: editor formatted (1) identically to author
%Control: production of article title (0) allowed
%Control: page (0) single
%Control: year (1) truncated
%Control: production of eprint (0) enabled
\begin{thebibliography}{31}%
\makeatletter
\providecommand \@ifxundefined [1]{%
 \@ifx{#1\undefined}
}%
\providecommand \@ifnum [1]{%
 \ifnum #1\expandafter \@firstoftwo
 \else \expandafter \@secondoftwo
 \fi
}%
\providecommand \@ifx [1]{%
 \ifx #1\expandafter \@firstoftwo
 \else \expandafter \@secondoftwo
 \fi
}%
\providecommand \natexlab [1]{#1}%
\providecommand \enquote  [1]{``#1''}%
\providecommand \bibnamefont  [1]{#1}%
\providecommand \bibfnamefont [1]{#1}%
\providecommand \citenamefont [1]{#1}%
\providecommand \href@noop [0]{\@secondoftwo}%
\providecommand \href [0]{\begingroup \@sanitize@url \@href}%
\providecommand \@href[1]{\@@startlink{#1}\@@href}%
\providecommand \@@href[1]{\endgroup#1\@@endlink}%
\providecommand \@sanitize@url [0]{\catcode `\\12\catcode `\$12\catcode `\&12\catcode `\#12\catcode `\^12\catcode `\_12\catcode `\%12\relax}%
\providecommand \@@startlink[1]{}%
\providecommand \@@endlink[0]{}%
\providecommand \url  [0]{\begingroup\@sanitize@url \@url }%
\providecommand \@url [1]{\endgroup\@href {#1}{\urlprefix }}%
\providecommand \urlprefix  [0]{URL }%
\providecommand \Eprint [0]{\href }%
\providecommand \doibase [0]{https://doi.org/}%
\providecommand \selectlanguage [0]{\@gobble}%
\providecommand \bibinfo  [0]{\@secondoftwo}%
\providecommand \bibfield  [0]{\@secondoftwo}%
\providecommand \translation [1]{[#1]}%
\providecommand \BibitemOpen [0]{}%
\providecommand \bibitemStop [0]{}%
\providecommand \bibitemNoStop [0]{.\EOS\space}%
\providecommand \EOS [0]{\spacefactor3000\relax}%
\providecommand \BibitemShut  [1]{\csname bibitem#1\endcsname}%
\let\auto@bib@innerbib\@empty
%</preamble>
\bibitem [{\citenamefont {Datta}\ and\ \citenamefont {Das}(1990)}]{Datta1990}%
  \BibitemOpen
  \bibfield  {author} {\bibinfo {author} {\bibfnamefont {S.}~\bibnamefont {Datta}}\ and\ \bibinfo {author} {\bibfnamefont {B.}~\bibnamefont {Das}},\ }\bibfield  {title} {\bibinfo {title} {{Electronic analog of the electro‐optic modulator}},\ }\href {https://doi.org/10.1063/1.102730} {\bibfield  {journal} {\bibinfo  {journal} {Applied Physics Letters}\ }\textbf {\bibinfo {volume} {56}},\ \bibinfo {pages} {665} (\bibinfo {year} {1990})}\BibitemShut {NoStop}%
\bibitem [{\citenamefont {Chuang}\ \emph {et~al.}(2015)\citenamefont {Chuang}, \citenamefont {Ho}, \citenamefont {Smith}, \citenamefont {Sfigakis}, \citenamefont {Pepper}, \citenamefont {Chen}, \citenamefont {Fan}, \citenamefont {Griffiths}, \citenamefont {Farrer}, \citenamefont {Beere}, \citenamefont {Jones}, \citenamefont {Ritchie},\ and\ \citenamefont {Chen}}]{Chuang2015}%
  \BibitemOpen
  \bibfield  {author} {\bibinfo {author} {\bibfnamefont {P.}~\bibnamefont {Chuang}}, \bibinfo {author} {\bibfnamefont {S.-C.}\ \bibnamefont {Ho}}, \bibinfo {author} {\bibfnamefont {L.~W.}\ \bibnamefont {Smith}}, \bibinfo {author} {\bibfnamefont {F.}~\bibnamefont {Sfigakis}}, \bibinfo {author} {\bibfnamefont {M.}~\bibnamefont {Pepper}}, \bibinfo {author} {\bibfnamefont {C.-H.}\ \bibnamefont {Chen}}, \bibinfo {author} {\bibfnamefont {J.-C.}\ \bibnamefont {Fan}}, \bibinfo {author} {\bibfnamefont {J.~P.}\ \bibnamefont {Griffiths}}, \bibinfo {author} {\bibfnamefont {I.}~\bibnamefont {Farrer}}, \bibinfo {author} {\bibfnamefont {H.~E.}\ \bibnamefont {Beere}}, \bibinfo {author} {\bibfnamefont {G.~A.~C.}\ \bibnamefont {Jones}}, \bibinfo {author} {\bibfnamefont {D.~A.}\ \bibnamefont {Ritchie}},\ and\ \bibinfo {author} {\bibfnamefont {T.-M.}\ \bibnamefont {Chen}},\ }\bibfield  {title} {\bibinfo {title} {{All-electric all-semiconductor spin field-effect transistors}},\ }\href {https://doi.org/10.1038/nnano.2014.296}
  {\bibfield  {journal} {\bibinfo  {journal} {Nature Nanotechnology}\ }\textbf {\bibinfo {volume} {10}},\ \bibinfo {pages} {35} (\bibinfo {year} {2015})}\BibitemShut {NoStop}%
\bibitem [{\citenamefont {Oreg}\ \emph {et~al.}(2010)\citenamefont {Oreg}, \citenamefont {Refael},\ and\ \citenamefont {von Oppen}}]{Oreg2010}%
  \BibitemOpen
  \bibfield  {author} {\bibinfo {author} {\bibfnamefont {Y.}~\bibnamefont {Oreg}}, \bibinfo {author} {\bibfnamefont {G.}~\bibnamefont {Refael}},\ and\ \bibinfo {author} {\bibfnamefont {F.}~\bibnamefont {von Oppen}},\ }\bibfield  {title} {\bibinfo {title} {{Helical Liquids and Majorana Bound States in Quantum Wires}},\ }\href {https://doi.org/10.1103/PhysRevLett.105.177002} {\bibfield  {journal} {\bibinfo  {journal} {Physical Review Letters}\ }\textbf {\bibinfo {volume} {105}},\ \bibinfo {pages} {177002} (\bibinfo {year} {2010})}\BibitemShut {NoStop}%
\bibitem [{\citenamefont {Lutchyn}\ \emph {et~al.}(2010)\citenamefont {Lutchyn}, \citenamefont {Sau},\ and\ \citenamefont {{Das Sarma}}}]{Lutchyn2010}%
  \BibitemOpen
  \bibfield  {author} {\bibinfo {author} {\bibfnamefont {R.~M.}\ \bibnamefont {Lutchyn}}, \bibinfo {author} {\bibfnamefont {J.~D.}\ \bibnamefont {Sau}},\ and\ \bibinfo {author} {\bibfnamefont {S.}~\bibnamefont {{Das Sarma}}},\ }\bibfield  {title} {\bibinfo {title} {{Majorana Fermions and a Topological Phase Transition in Semiconductor-Superconductor Heterostructures}},\ }\href {https://doi.org/10.1103/PhysRevLett.105.077001} {\bibfield  {journal} {\bibinfo  {journal} {Physical Review Letters}\ }\textbf {\bibinfo {volume} {105}},\ \bibinfo {pages} {077001} (\bibinfo {year} {2010})}\BibitemShut {NoStop}%
\bibitem [{\citenamefont {Liu}\ \emph {et~al.}(2023)\citenamefont {Liu}, \citenamefont {Gul}, \citenamefont {Holmes}, \citenamefont {Chen}, \citenamefont {Farrer}, \citenamefont {Ritchie},\ and\ \citenamefont {Pepper}}]{Liu2023}%
  \BibitemOpen
  \bibfield  {author} {\bibinfo {author} {\bibfnamefont {L.}~\bibnamefont {Liu}}, \bibinfo {author} {\bibfnamefont {Y.}~\bibnamefont {Gul}}, \bibinfo {author} {\bibfnamefont {S.~N.}\ \bibnamefont {Holmes}}, \bibinfo {author} {\bibfnamefont {C.}~\bibnamefont {Chen}}, \bibinfo {author} {\bibfnamefont {I.}~\bibnamefont {Farrer}}, \bibinfo {author} {\bibfnamefont {D.~A.}\ \bibnamefont {Ritchie}},\ and\ \bibinfo {author} {\bibfnamefont {M.}~\bibnamefont {Pepper}},\ }\bibfield  {title} {\bibinfo {title} {{Possible zero-magnetic field fractional quantization in In0.75Ga0.25As heterostructures}},\ }\href {https://doi.org/10.1063/5.0170273} {\bibfield  {journal} {\bibinfo  {journal} {Applied Physics Letters}\ }\textbf {\bibinfo {volume} {123}},\ \bibinfo {pages} {183502} (\bibinfo {year} {2023})}\BibitemShut {NoStop}%
\bibitem [{\citenamefont {Delfanazari}\ \emph {et~al.}(2024)\citenamefont {Delfanazari}, \citenamefont {Li}, \citenamefont {Xiong}, \citenamefont {Ma}, \citenamefont {Puddy}, \citenamefont {Yi}, \citenamefont {Farrer}, \citenamefont {Komori}, \citenamefont {Robinson}, \citenamefont {Serra}, \citenamefont {Ritchie}, \citenamefont {Kelly}, \citenamefont {Joyce},\ and\ \citenamefont {Smith}}]{Delfanazari2024}%
  \BibitemOpen
  \bibfield  {author} {\bibinfo {author} {\bibfnamefont {K.}~\bibnamefont {Delfanazari}}, \bibinfo {author} {\bibfnamefont {J.}~\bibnamefont {Li}}, \bibinfo {author} {\bibfnamefont {Y.}~\bibnamefont {Xiong}}, \bibinfo {author} {\bibfnamefont {P.}~\bibnamefont {Ma}}, \bibinfo {author} {\bibfnamefont {R.~K.}\ \bibnamefont {Puddy}}, \bibinfo {author} {\bibfnamefont {T.}~\bibnamefont {Yi}}, \bibinfo {author} {\bibfnamefont {I.}~\bibnamefont {Farrer}}, \bibinfo {author} {\bibfnamefont {S.}~\bibnamefont {Komori}}, \bibinfo {author} {\bibfnamefont {J.~W.~A.}\ \bibnamefont {Robinson}}, \bibinfo {author} {\bibfnamefont {L.}~\bibnamefont {Serra}}, \bibinfo {author} {\bibfnamefont {D.~A.}\ \bibnamefont {Ritchie}}, \bibinfo {author} {\bibfnamefont {M.~J.}\ \bibnamefont {Kelly}}, \bibinfo {author} {\bibfnamefont {H.~J.}\ \bibnamefont {Joyce}},\ and\ \bibinfo {author} {\bibfnamefont {C.~G.}\ \bibnamefont {Smith}},\ }\bibfield  {title} {\bibinfo {title} {{Quantized conductance in hybrid split-gate arrays of superconducting
  quantum point contacts with semiconducting two-dimensional electron systems}},\ }\href {https://doi.org/10.1103/PhysRevApplied.21.014051} {\bibfield  {journal} {\bibinfo  {journal} {Phys. Rev. Appl.}\ }\textbf {\bibinfo {volume} {21}},\ \bibinfo {pages} {14051} (\bibinfo {year} {2024})}\BibitemShut {NoStop}%
\bibitem [{\citenamefont {Aghaee}\ \emph {et~al.}(2023)\citenamefont {Aghaee}, \citenamefont {Akkala}, \citenamefont {Alam}, \citenamefont {Ali}, \citenamefont {{Alcaraz Ramirez}}, \citenamefont {Andrzejczuk}, \citenamefont {Antipov}, \citenamefont {Aseev}, \citenamefont {Astafev}, \citenamefont {Bauer}, \citenamefont {Becker}, \citenamefont {Boddapati}, \citenamefont {Boekhout}, \citenamefont {Bommer}, \citenamefont {Bosma}, \citenamefont {Bourdet}, \citenamefont {Boutin}, \citenamefont {Caroff}, \citenamefont {Casparis}, \citenamefont {Cassidy}, \citenamefont {Chatoor}, \citenamefont {Christensen}, \citenamefont {Clay}, \citenamefont {Cole}, \citenamefont {Corsetti}, \citenamefont {Cui}, \citenamefont {Dalampiras}, \citenamefont {Dokania}, \citenamefont {de~Lange}, \citenamefont {de~Moor}, \citenamefont {{Estrada Salda{\~{n}}a}}, \citenamefont {Fallahi}, \citenamefont {Fathabad}, \citenamefont {Gamble}, \citenamefont {Gardner}, \citenamefont {Govender}, \citenamefont {Griggio}, \citenamefont {Grigoryan},
  \citenamefont {Gronin}, \citenamefont {Gukelberger}, \citenamefont {Hansen}, \citenamefont {Heedt}, \citenamefont {{Herranz Zamorano}}, \citenamefont {Ho}, \citenamefont {Holgaard}, \citenamefont {Ingerslev}, \citenamefont {Johansson}, \citenamefont {Jones}, \citenamefont {Kallaher}, \citenamefont {Karimi}, \citenamefont {Karzig}, \citenamefont {King}, \citenamefont {Kloster}, \citenamefont {Knapp}, \citenamefont {Kocon}, \citenamefont {Koski}, \citenamefont {Kostamo}, \citenamefont {Krogstrup}, \citenamefont {Kumar}, \citenamefont {Laeven}, \citenamefont {Larsen}, \citenamefont {Li}, \citenamefont {Lindemann}, \citenamefont {Love}, \citenamefont {Lutchyn}, \citenamefont {Madsen}, \citenamefont {Manfra}, \citenamefont {Markussen}, \citenamefont {Martinez}, \citenamefont {McNeil}, \citenamefont {Memisevic}, \citenamefont {Morgan}, \citenamefont {Mullally}, \citenamefont {Nayak}, \citenamefont {Nielsen}, \citenamefont {Nielsen}, \citenamefont {Nijholt}, \citenamefont {Nurmohamed}, \citenamefont {O'Farrell},
  \citenamefont {Otani}, \citenamefont {Pauka}, \citenamefont {Petersson}, \citenamefont {Petit}, \citenamefont {Pikulin}, \citenamefont {Preiss}, \citenamefont {Quintero-Perez}, \citenamefont {Rajpalke}, \citenamefont {Rasmussen}, \citenamefont {Razmadze}, \citenamefont {Reentila}, \citenamefont {Reilly}, \citenamefont {Rouse}, \citenamefont {Sadovskyy}, \citenamefont {Sainiemi}, \citenamefont {Schreppler}, \citenamefont {Sidorkin}, \citenamefont {Singh}, \citenamefont {Singh}, \citenamefont {Sinha}, \citenamefont {Sohr}, \citenamefont {{Stankevi\ifmmode \checkc\else {\v{c}}\fi}}, \citenamefont {Stek}, \citenamefont {Suominen}, \citenamefont {Suter}, \citenamefont {Svidenko}, \citenamefont {Teicher}, \citenamefont {Temuerhan}, \citenamefont {Thiyagarajah}, \citenamefont {Tholapi}, \citenamefont {Thomas}, \citenamefont {Toomey}, \citenamefont {Upadhyay}, \citenamefont {Urban}, \citenamefont {Vaitiek\.{e}nas}, \citenamefont {{Van Hoogdalem}}, \citenamefont {{Van Woerkom}}, \citenamefont {Viazmitinov},
  \citenamefont {Vogel}, \citenamefont {Waddy}, \citenamefont {Watson}, \citenamefont {Weston}, \citenamefont {Winkler}, \citenamefont {Yang}, \citenamefont {Yau}, \citenamefont {Yi}, \citenamefont {Yucelen}, \citenamefont {Webster}, \citenamefont {Zeisel},\ and\ \citenamefont {Zhao}}]{PhysRevB.107.245423}%
  \BibitemOpen
  \bibfield  {author} {\bibinfo {author} {\bibfnamefont {M.}~\bibnamefont {Aghaee}}, \bibinfo {author} {\bibfnamefont {A.}~\bibnamefont {Akkala}}, \bibinfo {author} {\bibfnamefont {Z.}~\bibnamefont {Alam}}, \bibinfo {author} {\bibfnamefont {R.}~\bibnamefont {Ali}}, \bibinfo {author} {\bibfnamefont {A.}~\bibnamefont {{Alcaraz Ramirez}}}, \bibinfo {author} {\bibfnamefont {M.}~\bibnamefont {Andrzejczuk}}, \bibinfo {author} {\bibfnamefont {A.~E.}\ \bibnamefont {Antipov}}, \bibinfo {author} {\bibfnamefont {P.}~\bibnamefont {Aseev}}, \bibinfo {author} {\bibfnamefont {M.}~\bibnamefont {Astafev}}, \bibinfo {author} {\bibfnamefont {B.}~\bibnamefont {Bauer}}, \bibinfo {author} {\bibfnamefont {J.}~\bibnamefont {Becker}}, \bibinfo {author} {\bibfnamefont {S.}~\bibnamefont {Boddapati}}, \bibinfo {author} {\bibfnamefont {F.}~\bibnamefont {Boekhout}}, \bibinfo {author} {\bibfnamefont {J.}~\bibnamefont {Bommer}}, \bibinfo {author} {\bibfnamefont {T.}~\bibnamefont {Bosma}}, \bibinfo {author} {\bibfnamefont {L.}~\bibnamefont
  {Bourdet}}, \bibinfo {author} {\bibfnamefont {S.}~\bibnamefont {Boutin}}, \bibinfo {author} {\bibfnamefont {P.}~\bibnamefont {Caroff}}, \bibinfo {author} {\bibfnamefont {L.}~\bibnamefont {Casparis}}, \bibinfo {author} {\bibfnamefont {M.}~\bibnamefont {Cassidy}}, \bibinfo {author} {\bibfnamefont {S.}~\bibnamefont {Chatoor}}, \bibinfo {author} {\bibfnamefont {A.~W.}\ \bibnamefont {Christensen}}, \bibinfo {author} {\bibfnamefont {N.}~\bibnamefont {Clay}}, \bibinfo {author} {\bibfnamefont {W.~S.}\ \bibnamefont {Cole}}, \bibinfo {author} {\bibfnamefont {F.}~\bibnamefont {Corsetti}}, \bibinfo {author} {\bibfnamefont {A.}~\bibnamefont {Cui}}, \bibinfo {author} {\bibfnamefont {P.}~\bibnamefont {Dalampiras}}, \bibinfo {author} {\bibfnamefont {A.}~\bibnamefont {Dokania}}, \bibinfo {author} {\bibfnamefont {G.}~\bibnamefont {de~Lange}}, \bibinfo {author} {\bibfnamefont {M.}~\bibnamefont {de~Moor}}, \bibinfo {author} {\bibfnamefont {J.~C.}\ \bibnamefont {{Estrada Salda{\~{n}}a}}}, \bibinfo {author} {\bibfnamefont
  {S.}~\bibnamefont {Fallahi}}, \bibinfo {author} {\bibfnamefont {Z.~H.}\ \bibnamefont {Fathabad}}, \bibinfo {author} {\bibfnamefont {J.}~\bibnamefont {Gamble}}, \bibinfo {author} {\bibfnamefont {G.}~\bibnamefont {Gardner}}, \bibinfo {author} {\bibfnamefont {D.}~\bibnamefont {Govender}}, \bibinfo {author} {\bibfnamefont {F.}~\bibnamefont {Griggio}}, \bibinfo {author} {\bibfnamefont {R.}~\bibnamefont {Grigoryan}}, \bibinfo {author} {\bibfnamefont {S.}~\bibnamefont {Gronin}}, \bibinfo {author} {\bibfnamefont {J.}~\bibnamefont {Gukelberger}}, \bibinfo {author} {\bibfnamefont {E.~B.}\ \bibnamefont {Hansen}}, \bibinfo {author} {\bibfnamefont {S.}~\bibnamefont {Heedt}}, \bibinfo {author} {\bibfnamefont {J.}~\bibnamefont {{Herranz Zamorano}}}, \bibinfo {author} {\bibfnamefont {S.}~\bibnamefont {Ho}}, \bibinfo {author} {\bibfnamefont {U.~L.}\ \bibnamefont {Holgaard}}, \bibinfo {author} {\bibfnamefont {H.}~\bibnamefont {Ingerslev}}, \bibinfo {author} {\bibfnamefont {L.}~\bibnamefont {Johansson}}, \bibinfo {author}
  {\bibfnamefont {J.}~\bibnamefont {Jones}}, \bibinfo {author} {\bibfnamefont {R.}~\bibnamefont {Kallaher}}, \bibinfo {author} {\bibfnamefont {F.}~\bibnamefont {Karimi}}, \bibinfo {author} {\bibfnamefont {T.}~\bibnamefont {Karzig}}, \bibinfo {author} {\bibfnamefont {E.}~\bibnamefont {King}}, \bibinfo {author} {\bibfnamefont {M.~E.}\ \bibnamefont {Kloster}}, \bibinfo {author} {\bibfnamefont {C.}~\bibnamefont {Knapp}}, \bibinfo {author} {\bibfnamefont {D.}~\bibnamefont {Kocon}}, \bibinfo {author} {\bibfnamefont {J.}~\bibnamefont {Koski}}, \bibinfo {author} {\bibfnamefont {P.}~\bibnamefont {Kostamo}}, \bibinfo {author} {\bibfnamefont {P.}~\bibnamefont {Krogstrup}}, \bibinfo {author} {\bibfnamefont {M.}~\bibnamefont {Kumar}}, \bibinfo {author} {\bibfnamefont {T.}~\bibnamefont {Laeven}}, \bibinfo {author} {\bibfnamefont {T.}~\bibnamefont {Larsen}}, \bibinfo {author} {\bibfnamefont {K.}~\bibnamefont {Li}}, \bibinfo {author} {\bibfnamefont {T.}~\bibnamefont {Lindemann}}, \bibinfo {author} {\bibfnamefont
  {J.}~\bibnamefont {Love}}, \bibinfo {author} {\bibfnamefont {R.}~\bibnamefont {Lutchyn}}, \bibinfo {author} {\bibfnamefont {M.~H.}\ \bibnamefont {Madsen}}, \bibinfo {author} {\bibfnamefont {M.}~\bibnamefont {Manfra}}, \bibinfo {author} {\bibfnamefont {S.}~\bibnamefont {Markussen}}, \bibinfo {author} {\bibfnamefont {E.}~\bibnamefont {Martinez}}, \bibinfo {author} {\bibfnamefont {R.}~\bibnamefont {McNeil}}, \bibinfo {author} {\bibfnamefont {E.}~\bibnamefont {Memisevic}}, \bibinfo {author} {\bibfnamefont {T.}~\bibnamefont {Morgan}}, \bibinfo {author} {\bibfnamefont {A.}~\bibnamefont {Mullally}}, \bibinfo {author} {\bibfnamefont {C.}~\bibnamefont {Nayak}}, \bibinfo {author} {\bibfnamefont {J.}~\bibnamefont {Nielsen}}, \bibinfo {author} {\bibfnamefont {W.~H.~P.}\ \bibnamefont {Nielsen}}, \bibinfo {author} {\bibfnamefont {B.}~\bibnamefont {Nijholt}}, \bibinfo {author} {\bibfnamefont {A.}~\bibnamefont {Nurmohamed}}, \bibinfo {author} {\bibfnamefont {E.}~\bibnamefont {O'Farrell}}, \bibinfo {author} {\bibfnamefont
  {K.}~\bibnamefont {Otani}}, \bibinfo {author} {\bibfnamefont {S.}~\bibnamefont {Pauka}}, \bibinfo {author} {\bibfnamefont {K.}~\bibnamefont {Petersson}}, \bibinfo {author} {\bibfnamefont {L.}~\bibnamefont {Petit}}, \bibinfo {author} {\bibfnamefont {D.~I.}\ \bibnamefont {Pikulin}}, \bibinfo {author} {\bibfnamefont {F.}~\bibnamefont {Preiss}}, \bibinfo {author} {\bibfnamefont {M.}~\bibnamefont {Quintero-Perez}}, \bibinfo {author} {\bibfnamefont {M.}~\bibnamefont {Rajpalke}}, \bibinfo {author} {\bibfnamefont {K.}~\bibnamefont {Rasmussen}}, \bibinfo {author} {\bibfnamefont {D.}~\bibnamefont {Razmadze}}, \bibinfo {author} {\bibfnamefont {O.}~\bibnamefont {Reentila}}, \bibinfo {author} {\bibfnamefont {D.}~\bibnamefont {Reilly}}, \bibinfo {author} {\bibfnamefont {R.}~\bibnamefont {Rouse}}, \bibinfo {author} {\bibfnamefont {I.}~\bibnamefont {Sadovskyy}}, \bibinfo {author} {\bibfnamefont {L.}~\bibnamefont {Sainiemi}}, \bibinfo {author} {\bibfnamefont {S.}~\bibnamefont {Schreppler}}, \bibinfo {author} {\bibfnamefont
  {V.}~\bibnamefont {Sidorkin}}, \bibinfo {author} {\bibfnamefont {A.}~\bibnamefont {Singh}}, \bibinfo {author} {\bibfnamefont {S.}~\bibnamefont {Singh}}, \bibinfo {author} {\bibfnamefont {S.}~\bibnamefont {Sinha}}, \bibinfo {author} {\bibfnamefont {P.}~\bibnamefont {Sohr}}, \bibinfo {author} {\bibfnamefont {T.~c.~{\v{s}}.}\ \bibnamefont {{Stankevi\ifmmode \checkc\else {\v{c}}\fi}}}, \bibinfo {author} {\bibfnamefont {L.}~\bibnamefont {Stek}}, \bibinfo {author} {\bibfnamefont {H.}~\bibnamefont {Suominen}}, \bibinfo {author} {\bibfnamefont {J.}~\bibnamefont {Suter}}, \bibinfo {author} {\bibfnamefont {V.}~\bibnamefont {Svidenko}}, \bibinfo {author} {\bibfnamefont {S.}~\bibnamefont {Teicher}}, \bibinfo {author} {\bibfnamefont {M.}~\bibnamefont {Temuerhan}}, \bibinfo {author} {\bibfnamefont {N.}~\bibnamefont {Thiyagarajah}}, \bibinfo {author} {\bibfnamefont {R.}~\bibnamefont {Tholapi}}, \bibinfo {author} {\bibfnamefont {M.}~\bibnamefont {Thomas}}, \bibinfo {author} {\bibfnamefont {E.}~\bibnamefont {Toomey}},
  \bibinfo {author} {\bibfnamefont {S.}~\bibnamefont {Upadhyay}}, \bibinfo {author} {\bibfnamefont {I.}~\bibnamefont {Urban}}, \bibinfo {author} {\bibfnamefont {S.}~\bibnamefont {Vaitiek\.{e}nas}}, \bibinfo {author} {\bibfnamefont {K.}~\bibnamefont {{Van Hoogdalem}}}, \bibinfo {author} {\bibfnamefont {D.}~\bibnamefont {{Van Woerkom}}}, \bibinfo {author} {\bibfnamefont {D.~V.}\ \bibnamefont {Viazmitinov}}, \bibinfo {author} {\bibfnamefont {D.}~\bibnamefont {Vogel}}, \bibinfo {author} {\bibfnamefont {S.}~\bibnamefont {Waddy}}, \bibinfo {author} {\bibfnamefont {J.}~\bibnamefont {Watson}}, \bibinfo {author} {\bibfnamefont {J.}~\bibnamefont {Weston}}, \bibinfo {author} {\bibfnamefont {G.~W.}\ \bibnamefont {Winkler}}, \bibinfo {author} {\bibfnamefont {C.~K.}\ \bibnamefont {Yang}}, \bibinfo {author} {\bibfnamefont {S.}~\bibnamefont {Yau}}, \bibinfo {author} {\bibfnamefont {D.}~\bibnamefont {Yi}}, \bibinfo {author} {\bibfnamefont {E.}~\bibnamefont {Yucelen}}, \bibinfo {author} {\bibfnamefont {A.}~\bibnamefont
  {Webster}}, \bibinfo {author} {\bibfnamefont {R.}~\bibnamefont {Zeisel}},\ and\ \bibinfo {author} {\bibfnamefont {R.}~\bibnamefont {Zhao}},\ }\bibfield  {title} {\bibinfo {title} {{InAs-Al hybrid devices passing the topological gap protocol}},\ }\href {https://doi.org/10.1103/PhysRevB.107.245423} {\bibfield  {journal} {\bibinfo  {journal} {Phys. Rev. B}\ }\textbf {\bibinfo {volume} {107}},\ \bibinfo {pages} {245423} (\bibinfo {year} {2023})}\BibitemShut {NoStop}%
\bibitem [{\citenamefont {Dresselhaus}(1955)}]{Dresselhaus1955}%
  \BibitemOpen
  \bibfield  {author} {\bibinfo {author} {\bibfnamefont {G.}~\bibnamefont {Dresselhaus}},\ }\bibfield  {title} {\bibinfo {title} {{Spin-Orbit Coupling Effects in Zinc Blende Structures}},\ }\href {https://doi.org/10.1103/PhysRev.100.580} {\bibfield  {journal} {\bibinfo  {journal} {Physical Review}\ }\textbf {\bibinfo {volume} {100}},\ \bibinfo {pages} {580} (\bibinfo {year} {1955})}\BibitemShut {NoStop}%
\bibitem [{\citenamefont {Rashba}(1960)}]{Rashba1960}%
  \BibitemOpen
  \bibfield  {author} {\bibinfo {author} {\bibfnamefont {E.}~\bibnamefont {Rashba}},\ }\bibfield  {title} {\bibinfo {title} {{Properties of semiconductors with an extremum loop. I. Cyclotron and combinational Resonance in a magnetic field perpendicular to the plane of the loop}},\ }\href@noop {} {\bibfield  {journal} {\bibinfo  {journal} {Sov. Phys.-Solid State}\ }\textbf {\bibinfo {volume} {2}},\ \bibinfo {pages} {1109} (\bibinfo {year} {1960})}\BibitemShut {NoStop}%
\bibitem [{\citenamefont {Ganichev}\ \emph {et~al.}(2004)\citenamefont {Ganichev}, \citenamefont {Bel'kov}, \citenamefont {Golub}, \citenamefont {Ivchenko}, \citenamefont {Schneider}, \citenamefont {Giglberger}, \citenamefont {Eroms}, \citenamefont {{De Boeck}}, \citenamefont {Borghs}, \citenamefont {Wegscheider}, \citenamefont {Weiss},\ and\ \citenamefont {Prettl}}]{Ganichev2004}%
  \BibitemOpen
  \bibfield  {author} {\bibinfo {author} {\bibfnamefont {S.~D.}\ \bibnamefont {Ganichev}}, \bibinfo {author} {\bibfnamefont {V.~V.}\ \bibnamefont {Bel'kov}}, \bibinfo {author} {\bibfnamefont {L.~E.}\ \bibnamefont {Golub}}, \bibinfo {author} {\bibfnamefont {E.~L.}\ \bibnamefont {Ivchenko}}, \bibinfo {author} {\bibfnamefont {P.}~\bibnamefont {Schneider}}, \bibinfo {author} {\bibfnamefont {S.}~\bibnamefont {Giglberger}}, \bibinfo {author} {\bibfnamefont {J.}~\bibnamefont {Eroms}}, \bibinfo {author} {\bibfnamefont {J.}~\bibnamefont {{De Boeck}}}, \bibinfo {author} {\bibfnamefont {G.}~\bibnamefont {Borghs}}, \bibinfo {author} {\bibfnamefont {W.}~\bibnamefont {Wegscheider}}, \bibinfo {author} {\bibfnamefont {D.}~\bibnamefont {Weiss}},\ and\ \bibinfo {author} {\bibfnamefont {W.}~\bibnamefont {Prettl}},\ }\bibfield  {title} {\bibinfo {title} {{Experimental Separation of Rashba and Dresselhaus Spin Splittings in Semiconductor Quantum Wells}},\ }\href {https://doi.org/10.1103/PhysRevLett.92.256601} {\bibfield
  {journal} {\bibinfo  {journal} {Physical Review Letters}\ }\textbf {\bibinfo {volume} {92}},\ \bibinfo {pages} {256601} (\bibinfo {year} {2004})}\BibitemShut {NoStop}%
\bibitem [{\citenamefont {Giglberger}\ \emph {et~al.}(2007)\citenamefont {Giglberger}, \citenamefont {Golub}, \citenamefont {Bel'kov}, \citenamefont {Danilov}, \citenamefont {Schuh}, \citenamefont {Gerl}, \citenamefont {Rohlfing}, \citenamefont {Stahl}, \citenamefont {Wegscheider}, \citenamefont {Weiss}, \citenamefont {Prettl},\ and\ \citenamefont {Ganichev}}]{Giglberger2007}%
  \BibitemOpen
  \bibfield  {author} {\bibinfo {author} {\bibfnamefont {S.}~\bibnamefont {Giglberger}}, \bibinfo {author} {\bibfnamefont {L.~E.}\ \bibnamefont {Golub}}, \bibinfo {author} {\bibfnamefont {V.~V.}\ \bibnamefont {Bel'kov}}, \bibinfo {author} {\bibfnamefont {S.~N.}\ \bibnamefont {Danilov}}, \bibinfo {author} {\bibfnamefont {D.}~\bibnamefont {Schuh}}, \bibinfo {author} {\bibfnamefont {C.}~\bibnamefont {Gerl}}, \bibinfo {author} {\bibfnamefont {F.}~\bibnamefont {Rohlfing}}, \bibinfo {author} {\bibfnamefont {J.}~\bibnamefont {Stahl}}, \bibinfo {author} {\bibfnamefont {W.}~\bibnamefont {Wegscheider}}, \bibinfo {author} {\bibfnamefont {D.}~\bibnamefont {Weiss}}, \bibinfo {author} {\bibfnamefont {W.}~\bibnamefont {Prettl}},\ and\ \bibinfo {author} {\bibfnamefont {S.~D.}\ \bibnamefont {Ganichev}},\ }\bibfield  {title} {\bibinfo {title} {{Rashba and Dresselhaus spin splittings in semiconductor quantum wells measured by spin photocurrents}},\ }\href {https://doi.org/10.1103/PhysRevB.75.035327} {\bibfield  {journal}
  {\bibinfo  {journal} {Physical Review B}\ }\textbf {\bibinfo {volume} {75}},\ \bibinfo {pages} {035327} (\bibinfo {year} {2007})}\BibitemShut {NoStop}%
\bibitem [{\citenamefont {Knox}\ \emph {et~al.}(2018)\citenamefont {Knox}, \citenamefont {Li}, \citenamefont {Rosamond}, \citenamefont {Linfield},\ and\ \citenamefont {Marrows}}]{Knox2018}%
  \BibitemOpen
  \bibfield  {author} {\bibinfo {author} {\bibfnamefont {C.~S.}\ \bibnamefont {Knox}}, \bibinfo {author} {\bibfnamefont {L.~H.}\ \bibnamefont {Li}}, \bibinfo {author} {\bibfnamefont {M.~C.}\ \bibnamefont {Rosamond}}, \bibinfo {author} {\bibfnamefont {E.~H.}\ \bibnamefont {Linfield}},\ and\ \bibinfo {author} {\bibfnamefont {C.~H.}\ \bibnamefont {Marrows}},\ }\bibfield  {title} {\bibinfo {title} {{Deconvolution of Rashba and Dresselhaus spin-orbit coupling by crystal axis dependent measurements of coupled InAs/GaSb quantum wells}},\ }\href {https://doi.org/10.1103/PhysRevB.98.155323} {\bibfield  {journal} {\bibinfo  {journal} {Physical Review B}\ }\textbf {\bibinfo {volume} {98}},\ \bibinfo {pages} {155323} (\bibinfo {year} {2018})}\BibitemShut {NoStop}%
\bibitem [{\citenamefont {Farzaneh}\ \emph {et~al.}(2024)\citenamefont {Farzaneh}, \citenamefont {Hatefipour}, \citenamefont {Schiela}, \citenamefont {Lotfizadeh}, \citenamefont {Yu}, \citenamefont {Elfeky}, \citenamefont {Strickland}, \citenamefont {Matos-Abiague},\ and\ \citenamefont {Shabani}}]{Farzaneh2024}%
  \BibitemOpen
  \bibfield  {author} {\bibinfo {author} {\bibfnamefont {S.~M.}\ \bibnamefont {Farzaneh}}, \bibinfo {author} {\bibfnamefont {M.}~\bibnamefont {Hatefipour}}, \bibinfo {author} {\bibfnamefont {W.~F.}\ \bibnamefont {Schiela}}, \bibinfo {author} {\bibfnamefont {N.}~\bibnamefont {Lotfizadeh}}, \bibinfo {author} {\bibfnamefont {P.}~\bibnamefont {Yu}}, \bibinfo {author} {\bibfnamefont {B.~H.}\ \bibnamefont {Elfeky}}, \bibinfo {author} {\bibfnamefont {W.~M.}\ \bibnamefont {Strickland}}, \bibinfo {author} {\bibfnamefont {A.}~\bibnamefont {Matos-Abiague}},\ and\ \bibinfo {author} {\bibfnamefont {J.}~\bibnamefont {Shabani}},\ }\bibfield  {title} {\bibinfo {title} {{Observing magnetoanisotropic weak antilocalization in near-surface quantum wells}},\ }\href {https://doi.org/10.1103/PhysRevResearch.6.013039} {\bibfield  {journal} {\bibinfo  {journal} {Physical Review Research}\ }\textbf {\bibinfo {volume} {6}},\ \bibinfo {pages} {013039} (\bibinfo {year} {2024})}\BibitemShut {NoStop}%
\bibitem [{\citenamefont {Jana}\ and\ \citenamefont {Jena}(2011)}]{Jana2011}%
  \BibitemOpen
  \bibfield  {author} {\bibinfo {author} {\bibfnamefont {R.~K.}\ \bibnamefont {Jana}}\ and\ \bibinfo {author} {\bibfnamefont {D.}~\bibnamefont {Jena}},\ }\bibfield  {title} {\bibinfo {title} {{Stark-effect scattering in rough quantum wells}},\ }\href {https://doi.org/10.1063/1.3607485} {\bibfield  {journal} {\bibinfo  {journal} {Applied Physics Letters}\ }\textbf {\bibinfo {volume} {99}},\ \bibinfo {pages} {012104} (\bibinfo {year} {2011})}\BibitemShut {NoStop}%
\bibitem [{\citenamefont {Wickramasinghe}\ \emph {et~al.}(2018)\citenamefont {Wickramasinghe}, \citenamefont {Mayer}, \citenamefont {Yuan}, \citenamefont {Nguyen}, \citenamefont {Jiao}, \citenamefont {Manucharyan},\ and\ \citenamefont {Shabani}}]{Wickramasinghe2018}%
  \BibitemOpen
  \bibfield  {author} {\bibinfo {author} {\bibfnamefont {K.~S.}\ \bibnamefont {Wickramasinghe}}, \bibinfo {author} {\bibfnamefont {W.}~\bibnamefont {Mayer}}, \bibinfo {author} {\bibfnamefont {J.}~\bibnamefont {Yuan}}, \bibinfo {author} {\bibfnamefont {T.}~\bibnamefont {Nguyen}}, \bibinfo {author} {\bibfnamefont {L.}~\bibnamefont {Jiao}}, \bibinfo {author} {\bibfnamefont {V.}~\bibnamefont {Manucharyan}},\ and\ \bibinfo {author} {\bibfnamefont {J.}~\bibnamefont {Shabani}},\ }\bibfield  {title} {\bibinfo {title} {{Transport properties of near surface InAs two-dimensional heterostructures}},\ }\href {https://doi.org/10.1063/1.5050413} {\bibfield  {journal} {\bibinfo  {journal} {Applied Physics Letters}\ }\textbf {\bibinfo {volume} {113}},\ \bibinfo {pages} {262104} (\bibinfo {year} {2018})}\BibitemShut {NoStop}%
\bibitem [{\citenamefont {Zhang}\ \emph {et~al.}(2023)\citenamefont {Zhang}, \citenamefont {Lindemann}, \citenamefont {Gardner}, \citenamefont {Gronin}, \citenamefont {Wu},\ and\ \citenamefont {Manfra}}]{Zhang2023}%
  \BibitemOpen
  \bibfield  {author} {\bibinfo {author} {\bibfnamefont {T.}~\bibnamefont {Zhang}}, \bibinfo {author} {\bibfnamefont {T.}~\bibnamefont {Lindemann}}, \bibinfo {author} {\bibfnamefont {G.~C.}\ \bibnamefont {Gardner}}, \bibinfo {author} {\bibfnamefont {S.}~\bibnamefont {Gronin}}, \bibinfo {author} {\bibfnamefont {T.}~\bibnamefont {Wu}},\ and\ \bibinfo {author} {\bibfnamefont {M.~J.}\ \bibnamefont {Manfra}},\ }\bibfield  {title} {\bibinfo {title} {{Mobility exceeding 100 000 ${\mathrm{cm}}^{2}/\mathrm{V}$ s in modulation-doped shallow InAs quantum wells coupled to epitaxial aluminum}},\ }\href {https://doi.org/10.1103/PhysRevMaterials.7.056201} {\bibfield  {journal} {\bibinfo  {journal} {Phys. Rev. Mater.}\ }\textbf {\bibinfo {volume} {7}},\ \bibinfo {pages} {56201} (\bibinfo {year} {2023})}\BibitemShut {NoStop}%
\bibitem [{\citenamefont {Dong}\ \emph {et~al.}(2024)\citenamefont {Dong}, \citenamefont {Gul}, \citenamefont {Engel}, \citenamefont {van Schijndel}, \citenamefont {Dempsey}, \citenamefont {Pepper},\ and\ \citenamefont {Palmstr\o{}m}}]{Dong2024prm}%
  \BibitemOpen
  \bibfield  {author} {\bibinfo {author} {\bibfnamefont {J.~T.}\ \bibnamefont {Dong}}, \bibinfo {author} {\bibfnamefont {Y.}~\bibnamefont {Gul}}, \bibinfo {author} {\bibfnamefont {A.~N.}\ \bibnamefont {Engel}}, \bibinfo {author} {\bibfnamefont {T.~A.~J.}\ \bibnamefont {van Schijndel}}, \bibinfo {author} {\bibfnamefont {C.~P.}\ \bibnamefont {Dempsey}}, \bibinfo {author} {\bibfnamefont {M.}~\bibnamefont {Pepper}},\ and\ \bibinfo {author} {\bibfnamefont {C.~J.}\ \bibnamefont {Palmstr\o{}m}},\ }\bibfield  {title} {\bibinfo {title} {{Enhanced mobility of ternary InGaAs quantum wells through digital alloying}},\ }\href {https://doi.org/10.1103/PhysRevMaterials.8.064601} {\bibfield  {journal} {\bibinfo  {journal} {Phys. Rev. Mater.}\ }\textbf {\bibinfo {volume} {8}},\ \bibinfo {pages} {64601} (\bibinfo {year} {2024})}\BibitemShut {NoStop}%
\bibitem [{\citenamefont {Eldridge}\ \emph {et~al.}(2011)\citenamefont {Eldridge}, \citenamefont {H{\"{u}}bner}, \citenamefont {Oertel}, \citenamefont {Harley}, \citenamefont {Henini},\ and\ \citenamefont {Oestreich}}]{Eldridge2011}%
  \BibitemOpen
  \bibfield  {author} {\bibinfo {author} {\bibfnamefont {P.~S.}\ \bibnamefont {Eldridge}}, \bibinfo {author} {\bibfnamefont {J.}~\bibnamefont {H{\"{u}}bner}}, \bibinfo {author} {\bibfnamefont {S.}~\bibnamefont {Oertel}}, \bibinfo {author} {\bibfnamefont {R.~T.}\ \bibnamefont {Harley}}, \bibinfo {author} {\bibfnamefont {M.}~\bibnamefont {Henini}},\ and\ \bibinfo {author} {\bibfnamefont {M.}~\bibnamefont {Oestreich}},\ }\bibfield  {title} {\bibinfo {title} {{Spin-orbit fields in asymmetric (001)-oriented GaAs/Al${}_{x}$Ga${}_{1\ensuremath{-}x}$As quantum wells}},\ }\href {https://doi.org/10.1103/PhysRevB.83.041301} {\bibfield  {journal} {\bibinfo  {journal} {Phys. Rev. B}\ }\textbf {\bibinfo {volume} {83}},\ \bibinfo {pages} {41301} (\bibinfo {year} {2011})}\BibitemShut {NoStop}%
\bibitem [{\citenamefont {Pingenot}\ and\ \citenamefont {Mullen}(2011)}]{Pingenot2011}%
  \BibitemOpen
  \bibfield  {author} {\bibinfo {author} {\bibfnamefont {J.}~\bibnamefont {Pingenot}}\ and\ \bibinfo {author} {\bibfnamefont {K.}~\bibnamefont {Mullen}},\ }\bibfield  {title} {\bibinfo {title} {{Theoretical Comparison of Rashba Spin-Orbit Coupling in Digitally, Discretely, and Continuously Alloyed Nanostructures}},\ }\href@noop {} {\bibfield  {journal} {\bibinfo  {journal} {arXiv}\ ,\ \bibinfo {pages} {1105.1804}} (\bibinfo {year} {2011})}\BibitemShut {NoStop}%
\bibitem [{\citenamefont {Holmes}\ \emph {et~al.}(2008)\citenamefont {Holmes}, \citenamefont {Simmonds}, \citenamefont {Beere}, \citenamefont {Sfigakis}, \citenamefont {Farrer}, \citenamefont {Ritchie},\ and\ \citenamefont {Pepper}}]{Holmes2008}%
  \BibitemOpen
  \bibfield  {author} {\bibinfo {author} {\bibfnamefont {S.~N.}\ \bibnamefont {Holmes}}, \bibinfo {author} {\bibfnamefont {P.~J.}\ \bibnamefont {Simmonds}}, \bibinfo {author} {\bibfnamefont {H.~E.}\ \bibnamefont {Beere}}, \bibinfo {author} {\bibfnamefont {F.}~\bibnamefont {Sfigakis}}, \bibinfo {author} {\bibfnamefont {I.}~\bibnamefont {Farrer}}, \bibinfo {author} {\bibfnamefont {D.~A.}\ \bibnamefont {Ritchie}},\ and\ \bibinfo {author} {\bibfnamefont {M.}~\bibnamefont {Pepper}},\ }\bibfield  {title} {\bibinfo {title} {{Bychkov–Rashba dominated band structure in an In\textsubscript{0.75}Ga\textsubscript{0.25}As–In\textsubscript{0.75}Al\textsubscript{0.25} device with spin-split carrier densities of <10\textsuperscript{11} cm\textsuperscript{-2}}},\ }\href {https://doi.org/10.1088/0953-8984/20/47/472207} {\bibfield  {journal} {\bibinfo  {journal} {Journal of physics: Condensed Matter}\ }\textbf {\bibinfo {volume} {20}},\ \bibinfo {pages} {472207} (\bibinfo {year} {2008})}\BibitemShut {NoStop}%
\bibitem [{\citenamefont {Vurgaftman}\ \emph {et~al.}(2020)\citenamefont {Vurgaftman}, \citenamefont {Lumb},\ and\ \citenamefont {Meyer}}]{Vurgaftman2021}%
  \BibitemOpen
  \bibfield  {author} {\bibinfo {author} {\bibfnamefont {I.}~\bibnamefont {Vurgaftman}}, \bibinfo {author} {\bibfnamefont {M.~P.}\ \bibnamefont {Lumb}},\ and\ \bibinfo {author} {\bibfnamefont {J.~R.}\ \bibnamefont {Meyer}},\ }\href {https://doi.org/10.1093/oso/9780198767275.003.0001} {\emph {\bibinfo {title} {{Bands and Photons in III-V Semiconductor Quantum Structures}}}}\ (\bibinfo  {publisher} {Oxford University Press},\ \bibinfo {year} {2020})\ pp.\ \bibinfo {pages} {275--293}\BibitemShut {NoStop}%
\bibitem [{\citenamefont {St{\"{o}}rmer}\ \emph {et~al.}(1986)\citenamefont {St{\"{o}}rmer}, \citenamefont {Eisenstein}, \citenamefont {Gossard}, \citenamefont {Wiegmann},\ and\ \citenamefont {Baldwin}}]{Stormer1986}%
  \BibitemOpen
  \bibfield  {author} {\bibinfo {author} {\bibfnamefont {H.~L.}\ \bibnamefont {St{\"{o}}rmer}}, \bibinfo {author} {\bibfnamefont {J.~P.}\ \bibnamefont {Eisenstein}}, \bibinfo {author} {\bibfnamefont {A.~C.}\ \bibnamefont {Gossard}}, \bibinfo {author} {\bibfnamefont {W.}~\bibnamefont {Wiegmann}},\ and\ \bibinfo {author} {\bibfnamefont {K.}~\bibnamefont {Baldwin}},\ }\bibfield  {title} {\bibinfo {title} {{Quantization of the Hall Effect in an Anisotropic Three-Dimensional Electronic System}},\ }\href {https://doi.org/10.1103/PhysRevLett.56.85} {\bibfield  {journal} {\bibinfo  {journal} {Physical Review Letters}\ }\textbf {\bibinfo {volume} {56}},\ \bibinfo {pages} {85} (\bibinfo {year} {1986})}\BibitemShut {NoStop}%
\bibitem [{\citenamefont {Raikh}\ and\ \citenamefont {Shahbazyan}(1994)}]{Raikh1994}%
  \BibitemOpen
  \bibfield  {author} {\bibinfo {author} {\bibfnamefont {M.~E.}\ \bibnamefont {Raikh}}\ and\ \bibinfo {author} {\bibfnamefont {T.~V.}\ \bibnamefont {Shahbazyan}},\ }\bibfield  {title} {\bibinfo {title} {{Magnetointersubband oscillations of conductivity in a two-dimensional electronic system}},\ }\href {https://doi.org/10.1103/PhysRevB.49.5531} {\bibfield  {journal} {\bibinfo  {journal} {Phys. Rev. B}\ }\textbf {\bibinfo {volume} {49}},\ \bibinfo {pages} {5531} (\bibinfo {year} {1994})}\BibitemShut {NoStop}%
\bibitem [{\citenamefont {Sander}\ \emph {et~al.}(1998)\citenamefont {Sander}, \citenamefont {Holmes}, \citenamefont {Harris}, \citenamefont {Maude},\ and\ \citenamefont {Portal}}]{Sander1998}%
  \BibitemOpen
  \bibfield  {author} {\bibinfo {author} {\bibfnamefont {T.}~\bibnamefont {Sander}}, \bibinfo {author} {\bibfnamefont {S.}~\bibnamefont {Holmes}}, \bibinfo {author} {\bibfnamefont {J.}~\bibnamefont {Harris}}, \bibinfo {author} {\bibfnamefont {D.}~\bibnamefont {Maude}},\ and\ \bibinfo {author} {\bibfnamefont {J.}~\bibnamefont {Portal}},\ }\bibfield  {title} {\bibinfo {title} {{Determination of the phase of magneto-intersubband scattering oscillations in heterojunctions and quantum wells}},\ }\href {https://doi.org/10.1103/PhysRevB.58.13856} {\bibfield  {journal} {\bibinfo  {journal} {Physical Review B - Condensed Matter and Materials Physics}\ }\textbf {\bibinfo {volume} {58}},\ \bibinfo {pages} {13856} (\bibinfo {year} {1998})}\BibitemShut {NoStop}%
\bibitem [{\citenamefont {Gurevich}\ and\ \citenamefont {Firsov}(1961)}]{Gurevich1961}%
  \BibitemOpen
  \bibfield  {author} {\bibinfo {author} {\bibfnamefont {V.~L.}\ \bibnamefont {Gurevich}}\ and\ \bibinfo {author} {\bibfnamefont {Y.~A.}\ \bibnamefont {Firsov}},\ }\bibfield  {title} {\bibinfo {title} {{On the theory of the electrical conductivity of semiconductors in a magnetic field}},\ }\href@noop {} {\bibfield  {journal} {\bibinfo  {journal} {Soviet Phys. JETP}\ }\textbf {\bibinfo {volume} {13}},\ \bibinfo {pages} {137} (\bibinfo {year} {1961})}\BibitemShut {NoStop}%
\bibitem [{\citenamefont {Tsui}\ \emph {et~al.}(1980)\citenamefont {Tsui}, \citenamefont {Englert}, \citenamefont {Cho},\ and\ \citenamefont {Gossard}}]{Tsui1980}%
  \BibitemOpen
  \bibfield  {author} {\bibinfo {author} {\bibfnamefont {D.~C.}\ \bibnamefont {Tsui}}, \bibinfo {author} {\bibfnamefont {T.}~\bibnamefont {Englert}}, \bibinfo {author} {\bibfnamefont {A.~Y.}\ \bibnamefont {Cho}},\ and\ \bibinfo {author} {\bibfnamefont {A.~C.}\ \bibnamefont {Gossard}},\ }\bibfield  {title} {\bibinfo {title} {{Observation of Magnetophonon Resonances in a Two-Dimensional Electronic System}},\ }\href {https://doi.org/10.1103/PhysRevLett.44.341} {\bibfield  {journal} {\bibinfo  {journal} {Phys. Rev. Lett.}\ }\textbf {\bibinfo {volume} {44}},\ \bibinfo {pages} {341} (\bibinfo {year} {1980})}\BibitemShut {NoStop}%
\bibitem [{\citenamefont {Chen}\ \emph {et~al.}(2018)\citenamefont {Chen}, \citenamefont {Holmes}, \citenamefont {Farrer}, \citenamefont {Beere},\ and\ \citenamefont {Ritchie}}]{Chen2018}%
  \BibitemOpen
  \bibfield  {author} {\bibinfo {author} {\bibfnamefont {C.}~\bibnamefont {Chen}}, \bibinfo {author} {\bibfnamefont {S.~N.}\ \bibnamefont {Holmes}}, \bibinfo {author} {\bibfnamefont {I.}~\bibnamefont {Farrer}}, \bibinfo {author} {\bibfnamefont {H.~E.}\ \bibnamefont {Beere}},\ and\ \bibinfo {author} {\bibfnamefont {D.~A.}\ \bibnamefont {Ritchie}},\ }\bibfield  {title} {\bibinfo {title} {{High mobility In\textsubscript{0.75}Ga\textsubscript{0.25}As quantum wells in an InAs phonon lattice}},\ }\href {https://doi.org/10.1088/1361-648X/aaa947} {\bibfield  {journal} {\bibinfo  {journal} {Journal of Physics Condensed Matter}\ }\textbf {\bibinfo {volume} {30}},\ \bibinfo {pages} {105705} (\bibinfo {year} {2018})}\BibitemShut {NoStop}%
\bibitem [{\citenamefont {Minkov}\ \emph {et~al.}(2020)\citenamefont {Minkov}, \citenamefont {Rut}, \citenamefont {Sherstobitov}, \citenamefont {Dvoretski}, \citenamefont {Mikhailov}, \citenamefont {Solov'ev}, \citenamefont {Chernov}, \citenamefont {Ivanov},\ and\ \citenamefont {Germanenko}}]{Minkov2020}%
  \BibitemOpen
  \bibfield  {author} {\bibinfo {author} {\bibfnamefont {G.~M.}\ \bibnamefont {Minkov}}, \bibinfo {author} {\bibfnamefont {O.~E.}\ \bibnamefont {Rut}}, \bibinfo {author} {\bibfnamefont {A.~A.}\ \bibnamefont {Sherstobitov}}, \bibinfo {author} {\bibfnamefont {S.~A.}\ \bibnamefont {Dvoretski}}, \bibinfo {author} {\bibfnamefont {N.~N.}\ \bibnamefont {Mikhailov}}, \bibinfo {author} {\bibfnamefont {V.~A.}\ \bibnamefont {Solov'ev}}, \bibinfo {author} {\bibfnamefont {M.~Y.}\ \bibnamefont {Chernov}}, \bibinfo {author} {\bibfnamefont {S.~V.}\ \bibnamefont {Ivanov}},\ and\ \bibinfo {author} {\bibfnamefont {A.~V.}\ \bibnamefont {Germanenko}},\ }\bibfield  {title} {\bibinfo {title} {{Magneto-intersubband oscillations in two-dimensional systems with an energy spectrum split due to spin-orbit interaction}},\ }\href {https://doi.org/10.1103/PhysRevB.101.245303} {\bibfield  {journal} {\bibinfo  {journal} {Physical Review B}\ }\textbf {\bibinfo {volume} {101}},\ \bibinfo {pages} {245303} (\bibinfo {year} {2020})}\BibitemShut
  {NoStop}%
\bibitem [{\citenamefont {Engels}\ \emph {et~al.}(1997)\citenamefont {Engels}, \citenamefont {Lange}, \citenamefont {Sch{\"{a}}pers},\ and\ \citenamefont {L{\"{u}}th}}]{Engels1997}%
  \BibitemOpen
  \bibfield  {author} {\bibinfo {author} {\bibfnamefont {G.}~\bibnamefont {Engels}}, \bibinfo {author} {\bibfnamefont {J.}~\bibnamefont {Lange}}, \bibinfo {author} {\bibfnamefont {T.}~\bibnamefont {Sch{\"{a}}pers}},\ and\ \bibinfo {author} {\bibfnamefont {H.}~\bibnamefont {L{\"{u}}th}},\ }\bibfield  {title} {\bibinfo {title} {{Experimental and theoretical approach to spin splitting in modulation-dopeAs/InP quantum wells for B→0}},\ }\href {https://doi.org/10.1103/PhysRevB.55.R1958} {\bibfield  {journal} {\bibinfo  {journal} {Physical Review B - Condensed Matter and Materials Physics}\ }\textbf {\bibinfo {volume} {55}},\ \bibinfo {pages} {R1958} (\bibinfo {year} {1997})}\BibitemShut {NoStop}%
\bibitem [{\citenamefont {Chen}\ \emph {et~al.}(2015)\citenamefont {Chen}, \citenamefont {Farrer}, \citenamefont {Holmes}, \citenamefont {Sfigakis}, \citenamefont {Fletcher}, \citenamefont {Beere},\ and\ \citenamefont {Ritchie}}]{Chen2015}%
  \BibitemOpen
  \bibfield  {author} {\bibinfo {author} {\bibfnamefont {C.}~\bibnamefont {Chen}}, \bibinfo {author} {\bibfnamefont {I.}~\bibnamefont {Farrer}}, \bibinfo {author} {\bibfnamefont {S.~N.}\ \bibnamefont {Holmes}}, \bibinfo {author} {\bibfnamefont {F.}~\bibnamefont {Sfigakis}}, \bibinfo {author} {\bibfnamefont {M.~P.}\ \bibnamefont {Fletcher}}, \bibinfo {author} {\bibfnamefont {H.~E.}\ \bibnamefont {Beere}},\ and\ \bibinfo {author} {\bibfnamefont {D.~A.}\ \bibnamefont {Ritchie}},\ }\bibfield  {title} {\bibinfo {title} {{Growth variations and scattering mechanisms in metamorphic In\textsubscript{0.75}Ga\textsubscript{0.25}As/In\textsubscript{0.75}Al\textsubscript{0.25}As quantum wells grown by molecular beam epitaxy}},\ }\href {https://doi.org/10.1016/j.jcrysgro.2015.02.038} {\bibfield  {journal} {\bibinfo  {journal} {Journal of Crystal Growth}\ }\textbf {\bibinfo {volume} {425}},\ \bibinfo {pages} {70} (\bibinfo {year} {2015})}\BibitemShut {NoStop}%
\bibitem [{\citenamefont {Sch{\"{a}}pers}\ \emph {et~al.}(1998)\citenamefont {Sch{\"{a}}pers}, \citenamefont {Engels}, \citenamefont {Lange}, \citenamefont {Klocke}, \citenamefont {Hollfelder},\ and\ \citenamefont {L{\"{u}}th}}]{Schapers1998}%
  \BibitemOpen
  \bibfield  {author} {\bibinfo {author} {\bibfnamefont {T.}~\bibnamefont {Sch{\"{a}}pers}}, \bibinfo {author} {\bibfnamefont {G.}~\bibnamefont {Engels}}, \bibinfo {author} {\bibfnamefont {J.}~\bibnamefont {Lange}}, \bibinfo {author} {\bibfnamefont {T.}~\bibnamefont {Klocke}}, \bibinfo {author} {\bibfnamefont {M.}~\bibnamefont {Hollfelder}},\ and\ \bibinfo {author} {\bibfnamefont {H.}~\bibnamefont {L{\"{u}}th}},\ }\bibfield  {title} {\bibinfo {title} {{Effect of the heterointerface on the spin splitting in modulation doped InxGa1-xAs/InP quantum wells for B→0}},\ }\href {https://doi.org/10.1063/1.367192} {\bibfield  {journal} {\bibinfo  {journal} {Journal of Applied Physics}\ }\textbf {\bibinfo {volume} {83}},\ \bibinfo {pages} {4324} (\bibinfo {year} {1998})}\BibitemShut {NoStop}%
\end{thebibliography}%

\end{document}